\documentstyle[11pt,twoside]{article}

\begin{document}

\title{A PARTICLE THEORY OF THE CASIMIR EFFECT}

\author{{\bf Patrick
Suppes}\thanks{E-mail:suppes@ockham.stanford.edu}{\bf, Adonai
S. Sant'Anna}\thanks{Permanent address: Departamento de Matem\'atica,
UFPR, C.P. 19081, 81530-900, Curitiba, PR, Brazil. E-mail:
adonai@gauss.mat.ufpr.br}\\Ventura Hall, Stanford
University\\Stanford, California 94305-4115, USA\and {\bf J. Acacio de
Barros}\thanks{E-mail: acacio@fisica.ufjf.br}\\Departamento de
F\'{\i}sica, UFJF\\36036-330, Juiz de Fora, MG, Brazil.}

\date{ }

\maketitle

In previous works Suppes and de Barros used a pure particle model to
derive interference effects, where individual photons have
well-defined trajectories, and hence no wave properties. In the
present paper we extend that description to account for the Casimir
effect. We consider that the linear momentum $\sum\frac{1}{2}\hbar
{\bf k}$ of the vacuum state in quantum electrodynamics corresponds to
the linear momentum of virtual photons. The Casimir effect, in the
cases of two parallel plates and the solid ball, is explained in terms
of the pressure caused by the photons. Contrary to quantum
electrodynamics, we assume a finite number of virtual photons.\\\\Key
words: photon, trajectories, QED, Casimir effect, quantum vacuum.

\section{Introduction}

	Suppes and de Barros (1994a, 1994b, 1996) began a foundational
analysis on diffraction of light which formulated a probabilistic
theory of photons with well-defined photon trajectories and without
wave properties. The wave properties come from the expectation density
of the photons. The photons are also regarded as virtual, because they
are not directly observable, including their anihilation of each other
(see assumption (5) below). What can be detected is the interaction
with matter. The meaning of {\em virtual} used here is not the same as
in quantum electrodynamics (QED). In summary, our assumptions are:

\begin{itemize}
\item{Photons are emitted by harmonically oscillating sources;}
\item{They have definite trajectories;}
\item{They have a probability of being scattered at a slit;}
\item{Absorbers, like sources, are periodic;}
\item{Photons have positive and negative states ($+$-photons and $-$-photons)
which locally interfere, when being absorbed;}
\item{Photons change their states when reflected by a perfect conductor.}
\end{itemize}

	The expected density of $\pm$-photons emitted at $t$ in the
interval $dt$ is given by \begin{equation} s_{\pm}(t) =
\frac{A_{s}}{2}(1\pm\cos\omega t), \end{equation} where $\omega$ is
the frequency of a harmonically oscillating source, $A_{s}$ is a
constant determined by the source, and $t$ is time. If a photon is
emitted at $t'$, $0\leq t'\leq t$, then at time $t$ the photon has
traveled (with speed $c$) a distance $r$, where \begin{equation} t-t'
= \frac{r}{c}.  \end{equation} The conditional space-time expectation
density of $\pm$-photons for a spherically symmetric source with given
periodicity $\omega$ is: \begin{equation} h_{\pm}(t,r|\omega) =
\frac{A}{8\pi r^{2}}\left( 1 \pm \cos\omega\left( t -
\frac{r}{c}\right)\right),\label{hpm} \end{equation} where $A$ is a
real constant.

	The scalar field defined in terms of the expectation density
$h_{\pm}(t,r|\omega)$ is
\begin{equation}
{\cal E} = {\cal E}_{0}\frac{h_{+}-h_{-}}{\sqrt{h_{+} + h_{-}}},\label{field}
\end{equation}
where ${\cal E}_{0}$ is a scalar physical constant. Using (\ref{hpm}),
(\ref{field}) may be rewritten for a spherically symmetric source as:
\begin{equation}
{\cal E} = {\cal E}_{0}\sqrt{\frac{A}{4\pi r^{2}}}\cos\omega\left(t -
\frac{r}{c}\right).\label{newfield}
\end{equation}

	In the present paper we construct a corpuscular model for the
Casimir effect, following the ideas ever proposed by Suppes and de
Barros (1994a, 1994b, 1996). First, we study the case of the parallel
plates, and then the solid ball.

\section{Quantum Vacuum}

	A pure particle theory must postulate properties of the
quantum vacuum if standard field-theoretic methods of computing the
Casimir effect, and similar phenomena such as the Lamb shift, are to
be closely approximated.

	In QED the vacuum state has an energy $\frac{1}{2}\hbar\omega$
and a linear momentum $\frac{1}{2}\hbar {\bf k}$. We consider, as a
first postulate, that $\frac{1}{2}\hbar {\bf k}$ corresponds to the
linear momentum of one virtual photon. Our second postulate is that we
have a distribution function of $k$, $f(k)$, which is a probability
density for the distribution of photons with respect to $k$. Our third
postulate establishes that both $+$-photons and $-$-photons contribute
to the delivery of linear momentum on a reflective surface. So, the
conditional expectation density of photons, given $k$, that strike a
point on the conductor surface is \begin{equation} h(t,r_{S}|k) =
h_{+}(t,r_{S}|k) + h_{-}(t,r_{S}|k), \end{equation} where $r_{S}$ is a
surface point.

\section{Parallel Plates}

	We consider here the case of two perfectly conducting parallel
plates, standing face to face in vacuum at a distance $d$ much smaller
than their lateral extensions. It is well known that such plates
attract each other with a force per unit area (pressure) due to the
vacuum energy, as predicted by Casimir (1948), given by
\begin{equation} P = -\frac{\pi^{2}\hbar c}{240d^{4}}.\label{HBGC}
\end{equation} Usually, such an attraction is explained in terms of
the vacuum field. We use a random distribution of oscillating sources
of photons, in the vacuum, which do not interfere with each other, to
derive (\ref{HBGC}) in our conceptual framework.

	The photons outside the plates that strike such surfaces act
to push the plates together, while reflections of the photons confined
between the plates push them apart. This idea is proposed by Milonni,
Cook, and Goggin (1988) and also presented in (Milonni, 1994), but not
actually developed from a pure particle viewpoint.

	The photons that we are considering must satisfy a probability
density $f(k)\geq 0$. Rather than assume an explicit expression for
$f(k)$ (which requires some assumptions about the virtual photons), we
prefer to state some properties that $f(k)$ must satisfy:

\begin{description}
\item [\normalsize (i)]
{$\int_{0}^{\infty}\int_{0}^{\infty}\int_{0}^{\infty}f(k)dk_{x}dk_{y}dk_{z} =
1$, and the mean and the variance of $f(k)$ are finite;}
\item [\normalsize (ii)] {There exists a constant $H$ such that
$h(t,r_{S}|k)<H$.}
\item [\normalsize (iii)] {$h(t,r_{S}|k)f(k)|_{k=0} = 1$, and all derivatives
of this expression vanish at $k_{z} = 0$.}
\end{description}

	From (i)$\sim$(iii) and assumptions made earlier, we may
infer, contrary to a standard result of QED, that in our theory the
number of virtual photons is finite for any bounded region of
space-time. We also infer that $h(t,r_{S}|k)f(k)|_{k=\infty} = 0$,
which is intuitively an expected property of a cutoff function. We
note that $h(t,r_{S}|k)$ and $f(k)$ are not identified as specific
functions. We have generalized from standard cutoff functions, such as
an exponential function, to give reasonable sufficient conditions that
many different functions satisfy. We do not know enough about the
quantum vacuum to derive a particular choice.

	We divide the $xyz$ space into parallelepipeds of sides
$L_{x}$, $L_{y}$, and $L_{z}$, as in the usual description of QED. So,
all $k_{x}$, $k_{y}$, and $k_{z}$ must assume discrete values, as is
explained in the next paragraphs.

	We note that when reflected a photon changes its state from
positive to negative and vice versa. This single change for perfect
conductors implies that the defined scalar field, given by
(\ref{field}), vanishes at the reflecting surface. For further details
see (Suppes and de Barros, 1996). So, according to (\ref{newfield})
and recalling that $k = \omega/c$, we have at the wall:
\begin{equation} \cos \left(\omega t - \omega\frac{r}{c}\right) = \cos
\left(\omega t - kr\right) = 0.  \end{equation} If we set $\omega t =
\pi/2$, which corresponds to a convenient choice for the origin of
time, it is easy to see that the values of $k_{x}$, $k_{y}$, and
$k_{z}$ that satisfy the boundary condition in $x = L_{x}$, $y =
L_{y}$, and $z = L_{z}$ are: \begin{equation} \frac{k_{x,y,z}}{\pi} =
\frac{n}{L_{x,y,z}}.\label{periodic} \end{equation}

	But the condition that the scalar electric field vanishes at
the surface of the reflecting walls is not sufficient to explain the
periodicity given by (\ref{periodic}). A natural question arises: what
about the photons with linear momenta that do not satisfy
(\ref{periodic})? We recall that reflectors, like absorbers (Suppes
and de Barros, 1994b), behave periodically, since the photons are
continuously hitting the plates. Thus, the probability of reflecting a
photon is given by: \begin{equation} p = C(1+\cos(\omega t +
\psi)),\label{probability} \end{equation} where $\psi$ is a certain
phase. If $p = 0$, then there is no interaction with the plates, which
means that no momentum is delivered to it.

	As an example, consider the first strike of a photon on a
plate perpendicular to the $z$ axis. Such a surface is not oscillating
before the strike. But after reflection, the wall oscillates with the
same frequency $\omega$ associated to the linear momentum $k =
\omega/c$. The particle reflects on the other wall and returns to the
first wall with a phase $2L_{z}k_{z}$. But we must have $2L_{z}k_{z} =
2n\pi$, from (\ref{probability}), if the particle is to be reflected
again on its return to the first wall. Obviously, $\cos(\omega t -
2L_{z}k_{z}) = \cos(\omega t)$ if and only if $2L_{z}k_{z} = 2n\pi$.

	If we consider $L_{x,y,z}$ very large compared with any
physical dimensions of interest, we can assume that the $k_{x,y,z}$
approach a continuum. This is what holds for photons outside the
plates.

	Now we start to derive the pressure on the plates. We begin
with the inward pressure. The expected number of photons that strike
the area $dS$ of one of the plates, within the time interval $dt$ is
\begin{equation}
h(t,r_{S}|k)f(k)\frac{1}{\pi^{3}}dk_{x}dk_{y}dk_{z}\cos \gamma\,
c\,dt\,dS, \end{equation} where $\gamma$ is the angle of incidence of
the photons on the plate with respect to the normal of the surface,
i.e., $\cos\gamma = k_{z}/k$, where $k =
\sqrt{k_{x}^{2}+k_{y}^{2}+k_{z}^{2}}$. Thus, the element of volume
that we are taking into account is $\cos\gamma\, c\,dt\,dS$. The
factor $\frac{1}{\pi^{3}}$ is justified by (\ref{periodic}), since
outside the plates we approach the continuum as a limit.

	The momentum delivered to the plate by a single reflected
photon is equal to the negative of the change in the momentum of the
photon. In other words the momentum is equal to $2\frac{1}{2}\hbar
k_{z}$, if we consider the plate perpendicular to the $z$ component of
the $xyz$ system of coordinates. Therefore, the expected linear
momentum transfered to an area $dS$ on the plate during the time
interval $dt$ is \begin{equation}
\frac{\hbar}{\pi^{3}}\frac{k_{z}^{2}}{k}h(t,r_{S}|k)f(k)dk_{x}dk_{y}dk_{z}c\,dt\,dS.\label{momentum}
\end{equation}

	The force on the plate is obtained by dividing
(\ref{momentum}) by $dt$. The pressure is obtained by dividing the
force by $dS$. We denote the inward pressure as $P_{in}$ and the
outward pressure as $P_{out}$. Hence: \begin{equation} dP_{in} =
\frac{\hbar c}{\pi^{3}}\frac{
k_{z}^{2}h(t,r_{S}|k)f(k)}{\sqrt{k_{x}^{2}+k_{y}^{2}+k_{z}^{2}}}
dk_{x}dk_{y}dk_{z}.  \end{equation}
	Integrating over momentum: \begin{equation} P_{in} =
\frac{\hbar
c}{\pi^{3}}\int_{0}^{\infty}dk_{x}\int_{0}^{\infty}dk_{y}\int_{0}^{\infty}dk_{z}\frac{h(t,r_{S}|k)f(k)k_{z}^{2}}{\sqrt{k_{x}^{2}+k_{y}^{2}+k_{z}^{2}}}.
\end{equation} The equation given above is identical to a result due
to Milonni, Cook and Goggin (1988), if we consider that
$h(t,r_{S}|k)f(k)$ has the role of the usual cutoff function.

	To obtain the expression of the outward pressure we use
similar arguments. But now, because of the small distance $d$ between
the plates, we must take into account the periodicity given in
(\ref{periodic}), at least for the $z$ component. Hence:
\begin{equation} P_{out} = \frac{\hbar
c}{\pi^{2}d}\sum_{n=1}^{\infty}\int_{0}^{\infty}dk_{x}\int_{0}^{\infty}dk_{y}\frac{h(t,r_{S}|k)f(k)(\frac{n\pi}{d})^{2}}{\sqrt{k_{x}^{2}+k_{y}^{2}+(\frac{n\pi}{d})^{2}}}.
\end{equation} We note that it follows from (i)$\sim$(iii) that
$P_{in}$ and $P_{out}$ are both finite.

	The resultant pressure is given by:

\[P_{out} - P_{in} = \] \[\frac{\pi^{2}\hbar
c}{4d^{4}}\sum_{n=1}^{\infty}n^{2}
\int_{0}^{\infty}dx\frac{h(t,r_{S}|x,u)f(\sqrt{x+n^{2}})}{\sqrt{x+n^{2}}}
-\] \begin{equation} \frac{\pi^{2}\hbar
c}{4d^{4}}\int_{0}^{\infty}du\,u^{2}\int_{0}^{\infty}dx\frac{h(t,r_{S}|x,u)f(\sqrt{x+u^{2}})}{\sqrt{x+u^{2}}},\label{diff}
\end{equation} where, by change of variables, $f(k) =
f(\sqrt{x+u^{2}})$, $x = x'^{2} =
\frac{k_{x}^{2}d^{2}}{\pi^{2}}+\frac{k_{y}^{2}d^{2}}{\pi^{2}}$, $u =
k_{z}\frac{d}{\pi}$, $\theta = \tan\left(\frac{k_{y}}{k_{x}}\right)$,
and $dk_{x}dk_{y} = x'dx'd\theta\frac{\pi^{2}}{d^{2}}$. The expression
$h(t,r_{S}|x,u)f(\sqrt{x+u^{2}})$ corresponds to a cutoff function. In
our model $h(t,r_{S}|k)$ is bounded and $f(\sqrt{x+u^{2}})$ has the
physical interpretation of a probability density of the frequencies of
the photons.

	Frequently it is assumed that the cutoff function has the
property of going to zero as $k$ approaches infinity and going to one
when $k$ approaches zero. This is justified physically with the
hypothesis that the conductivity of the reflecting conductors
decreases to zero as the frequency gets high. Since $h(t,r_{S}|k)$ is
bounded, it is easy to see that the product $h(t,r_{S}|k)f(k)$ must
assume a similar role with respect to the cutoff, from a mathematical
standpoint. According to the assumptions that we made about
$h(t,r_{S}|k)$ and $f(k)$, all the properties of a cutoff function are
satisfied for $h(t,r_{S}|k)f(k)$.

	If we consider: \begin{equation} F(u) =
u^{2}\int_{0}^{\infty}dx\frac{h(t,r_{S}|x,u)f(\sqrt{x+u^{2}})}{\sqrt{x+u^{2}}},\label{Fu}
\end{equation} it is clear that the Euler-MacLaurin summation formula
(Abramowitz and Stegun, 1971) may be applied to (\ref{diff}). Contrary
to the standard QED treatment, we apply this formula to a convergent
rather than a divergent series. Therefore, the factor that is
multiplying $\frac{\pi^{2}\hbar c}{4d^{4}}$ in (\ref{diff}) may be
written as: \begin{equation} \sum_{n=1}^{\infty}F(n) -
\int_{0}^{\infty}duF(u) = -\frac{1}{2}F(0) - \frac{1}{12}F'(0) +
\frac{1}{720}F'''(0)...  \end{equation} for
$\lim_{u\rightarrow\infty}F(u) = 0$, since $\sum_{n=1}^{\infty}F(n)$
is finite and so $\lim_{n\rightarrow\infty}F(n) = 0$. We note that
$F(0) = 0$, $F'(0) = 0$, $F'''(0) = -12h(t,r_{S}|0)$, and all higher
derivatives $F^{(n)}(0)$, where $n$ is odd, vanish in accordance with
assumption (iii), as is shown in the appendix. Since by (iii)
$h(t,r_{S}|0)f(0) = 1$: \begin{equation} P_{out} - P_{in} =
-\frac{\pi^{2}\hbar c}{240d^{4}}.\label{HBGCp} \end{equation}

	Equation (\ref{HBGCp}) is identical to (\ref{HBGC}), which
completes our derivation of the Casimir effect for parallel plates.

\section{The Solid Ball}

	Many authors have considered the case of the Casimir force for
solid balls and cavities like spheres, hemispheres, and
spheroids. Balian and Duplantier (1977) developed a method that
establishes an expansion for the Green functions describing
electromagnetic waves in the presence of a perfectly conducting
boundary. Later, they applied their method to the study of the Casimir
free energy of the electromagnetic field in regions bounded by thin
perfect conductors with arbitrary shape (Balian and Duplantier,
1978). Brevik and Einevoll (1988) used Schwinger's source theory to
establish the Casimir surface force in the case of a solid ball,
considering $\epsilon(\omega)\mu(\omega) = 1$, where
$\epsilon(\omega)$ is the spectral permittivity and $\mu(\omega)$ is
the spectral permeability. In 1990, Brevik and Sollie (1990)
calculated the Casimir surface force on a spherical shell, assuming
the same condition $\epsilon(\omega)\mu(\omega) = 1$. Barton (1991a,
1991b) uses standard statistical and quantum physics to analyze the
fluctuations of the Casimir stress exerted on a flat perfect conductor
by the vacuum electromagnetic fields in adjacent space. Eberlein
(1992) makes an extension of Barton's work, calculating the
mean-square forces acting on spheres and hemispheres of variable
sizes.

	We adopt here the same corpuscular model presented in last
section, to the case of a solid ball of radius $a$ surrounded by
vacuum. Consider a ball, centered at the origin of a spherical
coordinate system $(\rho,\varphi,\theta)$. At each point of the
surface of the ball, we define a Cartesian system of coordinates, with
axes $\bot$, $\Vert_{1}$, and $\Vert_{2}$. $\bot$ is the normal to the
surface, and $\Vert_{1}$ and $\Vert_{2}$ are tangent to the sphere.

	As in the case of the parallel plates, we assume a
distribution function $f(k)$ satisfying the same properties assumed in
the last section. The expected number of photons that strike the area
$dS = a^{2}\sin\varphi\, d\varphi\, d\theta$ on the surface of the
solid ball, within the time interval $dt$ is \begin{equation}
h(t,r_{S}|k)f(k)dk_{\bot}dk_{\Vert_{1}}dk_{\Vert_{2}}c\,dt\,\cos\gamma\,a^{2}\sin\varphi\,d\varphi\,d\theta.
\end{equation} where $\gamma$ is an angle of incidence of the photons
with respect to the normal of the surface, i.e., $\cos\gamma =
k_{\bot}/k$, and $k =
\sqrt{k_{\bot}^{2}+k_{\Vert_{1}}^{2}+k_{\Vert_{2}}^{2}}$.

	As in the case of the plates, the momentum delivered to the
ball by a single reflecting photon is $2\frac{1}{2}\hbar
k_{\bot}$. The linear momentum on the ball is \begin{equation}
\frac{\hbar
k_{\bot}^{2}h(t,r_{S}|k)f(k)}{\pi^{3}k}dk_{\bot}dk_{\Vert_{1}}dk_{\Vert_{2}}c\,dt\,\,a^{2}\sin\varphi\,d\varphi\,d\theta.
\end{equation} The force is obtained by dividing the expression above
by $dt$: \begin{equation} dF = \frac{\hbar
c}{\pi^{3}}a^{2}h(t,r_{S}|k)f(k)\frac{k_{\bot}^{2}}{\sqrt{k_{\bot}^{2}+k_{\Vert_{1}}^{2}+k_{\Vert_{2}}^{2}}}dk_{\bot}dk_{\Vert_{1}}dk_{\Vert_{2}}\sin\varphi\,d\varphi\,d\theta.\label{
} \end{equation} Integrating: \begin{equation} F = \frac{4a^{2}\hbar
c}{\pi^{2}}\int_{0}^{\infty}dk_{\bot}\int_{0}^{\infty}dk_{\Vert_{1}}\int_{0}^{\infty}dk_{\Vert_{2}}\frac{h(t,r_{S}|k)f(k)k_{\bot}^{2}}{\sqrt{k_{\bot}^{2}+k_{\Vert_{1}}^{2}+k_{\Vert_{2}}^{2}}}.\label{Milonni-2}
\end{equation} By the arguments used earlier, the force $F$ is finite.

	Our result depends explicitly on $h(t,r_{S}|k)f(k)$ in
(\ref{Milonni-2}), which has, as in the case of the parallel plates, a
role similar to a cutoff. This is a consequence of the geometry of the
problem. Brevik and Einevoll (1988) obtained another expression for
the Casimir force in the case of a solid ball, which directly depends
on a typical value ($3\times 10^{6} {\rm sec}^{-1}$) for the cutoff
frequency.

\section{Acknowledgments}

	We thank Pierre Noyes, Max Dresden and Yair Guttmann for
extensive comments and criticisms in the context of the Stanford
seminar on foundations of statistical mechanics. A.S.S. acknowledges
financial support from CNPq (Conselho Nacional de Desenvolvimento
Cient\'{\i}fico e Tecnol\'ogico, Brazil).

\appendix

\renewcommand{\thesection}{Appendix}

\section{ }

	It follows from (\ref{Fu}) that
\begin{equation}
F(u) = u^{2}\int_{0}^{\infty}dx\frac{g(\sqrt{x+u^{2}})}{\sqrt{x+u^{2}}} =
2u^{2}\int_{0}^{\infty}dx\,g(\sqrt{x+u^{2}})\frac{1}{2}(x+u^{2})^{-1/2},\label{app1}
\end{equation}
where $g(\sqrt{x+u^{2}}) = h(t,r_{S}|x,u)f(\sqrt{x+u^{2}})$. If we make $y =
(x+u^{2})^{1/2}$:
\begin{equation}
F(u) = 2u^{2}\int_{u}^{\infty}dy\,g(y).\label{app2}
\end{equation}

	Before the evaluation of the derivatives of $F(u)$, we must
observe that
\begin{equation}
\frac{d}{du}\int_{u}^{\infty}dy\,g(y) = -g(u),\label{app3}
\end{equation}
since $g(\infty) = 0$. Hence:
\begin{equation}
F'(u) = 4u\int_{u}^{\infty}dy\,g(y) - 2u^{2}g(u),\label{app4}
\end{equation}
\begin{equation}
F'''(u) = -12g(u) - 12ug'(u) -
2u^{2}g''(u),\label{app5}
\end{equation}
and all higher derivatives $F^{(2n+1)}(u)$ vanish at $u=0$ if the even
derivatives of $g(u)$ vanish at the same point, which is assumed in
(iii).

\end{document}